\documentclass[iop,revtex4]{emulateapj}
\usepackage{lineno}

\shorttitle{Speckle interferometry at SOAR}
\shortauthors{Tokovinin}

\begin{document}

\title{Ten years of speckle interferometry at SOAR}

\author{Andrei Tokovinin}

\email{atokovinin@ctio.noao.edu}

\affil{Cerro Tololo Inter-American Observatory, Casilla 603, La
  Serena, Chile}

\begin{abstract}
Since 2007,  close binary and  multiple stars are observed  by speckle
interferometry  at the  4.1 m  Southern Astrophysical  Research (SOAR)
telescope.  The  HRCam instrument, observing  strategy and planning,
data processing and calibration methods, developed and improved during
ten years, are  presented here in a concise  way.  Thousands of binary
stars  were measured with  diffraction-limited resolution  (29\,mas at
540\,nm wavelength)  and a high accuracy reaching  1\,mas; two hundred
new pairs or subsystems were discovered.  To date, HRCam has performed
over 11\,000 observations with a  high efficiency (up to 300 stars per
night).  An overview of the  main results delivered by this instrument
is given.
\end{abstract}



\section{Introduction}
\label{sec:intro}

Speckle   interferometry  (SI)  invented   by  \citet{Lab1970}   is  a
well-known method of  reaching diffraction-limited resolution at large
telescopes,  despite seeing  and other  distortions.  It  exploits the
fine structure of short-exposure  images caused by the interference of
light.  The same goal is achieved nowadays by means of adaptive optics
(AO). However, AO  works mostly in the infra-red  domain (hence with a
lower resolution); it is more complex, while typical overheads make it
less  efficient than  the SI  or its  flavor called  ``lucky imaging''
(LI).   However,  both   AO   and  SI/LI   can  work   simultaneously,
complementing and enhancing each other. 

Early SI instruments built in  the 1970s and 1980s employed electronic
image intensifiers  in conjunction with photographic  film and, later,
CCDs.   Development of  Electron-Multiplication (EM)  CCDs  capable of
detecting  single  photo-electrons  enabled  a  new,  more  performant
generation of SI cameras.  The High-Resolution Camera (HRCam) built in
2007  \citep{HRCam}  is one  of  the  first  such instruments.   Other
similar instruments are AstraLux \citep{AstraLux,Hippler2009}, the BTA
speckle   camera    \citep{BTA},   DSSI   \citep{Horch2012},   Robo-AO
\citep{Baranec2014}, AOLI \citep{AOLI}.

Current  proliferation of  SI/LI instruments  is motivated  by several
science  drivers.   Discovery of  exo-planets  attracted attention  to
nearby  stars and their  resolved companions.   On one  hand, binaries
perturb both photometry and radial velocities, forcing most exo-planet
surveys  to screen their  targets with  high spatial  resolution.  For
example,  follow-up of the  {\it Kepler}  objects is  one of  the main
objectives of  the speckle  program at the  Gemini telescopes.  On the
other hand,  the need to understand  the common origin  of stellar and
planetary  systems  has  stimulated  statistical  surveys  of  stellar
multiplicity,  where   the  SI/LI  became   the  enabling  technology.
Finally, the era  of precise astrometry opened by  {\it Hipparcos} and
{\it Gaia} requires ground-based support to disentangle orbital motion
of binaries  in the astrometric  reductions and to extend  the limited
temporal  coverage of these  missions.  The  science drivers  call for
observations  of   many  hundreds  or  thousands   of  targets.   The
efficiency  of SI  largely surpasses  that of  typical  AO instruments
\citep[note however][]{Riddle2015}  for a number of  reasons (e.g. the
need to acquire a guide star for AO), making it the method of choice.

The HRCam has been originally  described by \citet{HRCam}. The goal of
this paper is  to present this instrument and  its subsequent upgrades
in a more  complete and systematic way.  This  includes the methods of
data  reduction,  performance metrics,  and  the observing  procedure,
which  evolved during  ten years,  reflecting the  growing experience.
  To date, a  large number of binary
star measurements and discoveries resulted from observations made with
HRCam, justifying  detailed description of the  instrument and its
limitations in the present paper.

\begin{figure}[ht]
\epsscale{1.0}
\plotone{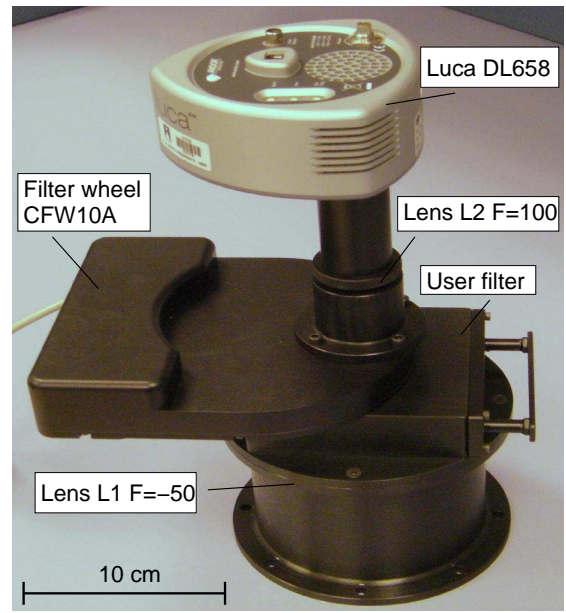}
\caption{HRCam and its main elements.
\label{fig:instrument}
}
\end{figure}

Section~\ref{sec:instr} describes the HRCam. Observations and data
reduction are covered in \S~\ref{sec:obs}, and some results from this
instrument are presented in \S~\ref{sec:res}. The paper closes with a
short summary in \S~\ref{sec:sum}. 


\section{High-Resolution Camera (HRCAM)}
\label{sec:instr}

\subsection{Optics and mechanics}
\label{sec:opt}

HRCam  works  at the  4.1  m  Southern  Astrophysical Research  (SOAR)
telescope \citep{SOAR}.  It was originally designed to help commission
the SOAR Adaptive Module (SAM) in its initial configuration, where the
natural guide stars  were used \citep{SAM}.  The choice  of the EM CCD
with 10 micron pixels called  for the magnification of the F/16.5 SOAR
focal  plane  by  two  times  to ensure  proper  Nyquist  sampling  of
speckles.  This is done by the combination of two achromatic lenses, a
negative  L1 with  $F =  -50$\,mm located  in front  of the  focus and
collimating the beam, and a positive  lens L2 with $F = +100$\,mm that
refocuses the magnified image on  the detector.  This simple optics is
diffraction-limited.  Later  HRCam worked with two other  EM CCDs with
smaller and larger pixels. In those cases, the L2 lens was replaced to
approximately preserve the image scale of 15\,mas per pixel.

Figure~\ref{fig:instrument}   shows   the   HRCam  in   its   original
configuration  with  the  Luca  EM CCD camera  (see  \S~\ref{sec:det}).   Its
commercial  components   are  listed  in   Table~\ref{tab:comp}.   The
mechanical design by P.~Schurter is very simple and modular. The lower
cylinder contains the negative lens L1.  A box with a sliding frame is
provided to hold a large user-defined filter; however, this option was
used only for technical work  with SAM. The commercial filter wheel is
attached to the box and, in turn, holds the tube with L2, to which the
camera  is attached  by its  C-mount thread.   The camera  and  L2 are
mutually focused to  infinity by adjusting the tube  length.  The tube
is  replaced  when  an  L2  lens  with a  different  focal  length  is
installed.  Unfortunately, this mechanics allows axial rotation of the
camera.   Therefore, its  position angle  has to  be adjusted  at each
re-installation.

\begin{table}[ht]
\center
\caption{Components of HRCam}
\label{tab:comp}
\begin{tabular}{l c l}
\hline
Element & Model & Vendor \\
\hline
Filter wheel & CFW10-SA & sbig.com \\
Filters,  1.25$''$ diam.   & $B,V,R,I$, H$\alpha$ & sbig.com \\
Filter $y$     & 543/22nm, \#76-032 & edmundoptics.com \\
Negative lens & $-$50\,mm,  \#62-492  & edmundoptics.com \\
Positive lens &  100\,mm, \#47-641  & edmundoptics.com \\
\hline
\end{tabular}
\end{table}

\begin{figure}[ht]
\epsscale{1.0}
\plotone{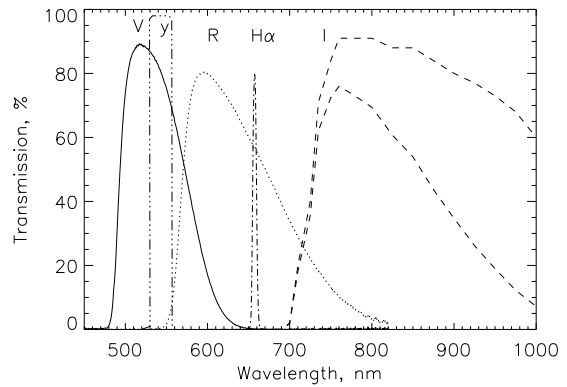}
\caption{Transmission curves of the HRCam filters. For the $I$ filter,
  the lower curve is the product of the filter transmission and the QE of the
  iXon-888 CCD; the latter defines the cutoff at long wavelengths.
\label{fig:filters}
}
\end{figure}

HRCam has a set of standard  $B,V R,I$ filters of $1.25''$ diameter in
its filter  wheel, as  well as the  narrow-band H$\alpha$  filter with
5\,nm bandwidth  (Figure~\ref{fig:filters}).  The originally installed
Str\"omgren $y$ filter  was cut out from the old  interference filter with
the  central   wavelength  551\,nm,  bandwidth   22\,nm,  and  maximum
transmission  of  57\%.   It  was  later replaced  by  the  commercial
interference filter  with a rectangular pass-band of  543/22\,nm and an
excellent transmission.

In 2017, the  Luca camera was replaced by  the more powerful detector,
iXon-888 (see \S~\ref{sec:det} and Table~\ref{tab:det}).  This heavier
camera required reinforcement of  the mechanical structure, because it
could no  longer be held by  the tube bolted  to the thin wall  of the
filter wheel. A plate supported  by the truss was designed to transfer
the load of  the camera directly to the mounting  plate of HRCam.  The
camera is still  connected to the tube, but is  also firmly clamped to
the new plate. The iXon-888 uses L2 with $F = 125$\,mm to preserve the
pixel scale.

HRCam is normally attached to the SAM as a user instrument. SAM relays
the image without change of the plate scale, optionally correcting the
seeing  by  its deformable  mirror  (DM). In  most  cases  the DM  was
passively  flattened   during  speckle  observations.    However,  for
observations  of  faint  targets,  the  AO compensation  was  used  to
concentrate the  light and thus to increase  the sensitivity.  Another
important  function of SAM  is to  correct the  atmospheric dispersion
(AD);  the AD corrector is  described  by  \citet{ADC}.   As SAM  became
available  only in 2009,  previous observations  with HRCam  were made
without  AD  correction.   In  this  case, the speckle  elongation  was
accounted for in the data reduction.

The guide probe  of SAM, located at the  original (uncorrected) focus,
can project  a point  source into the  instrument. This  capability is
used to control  the optical quality and to  calibrate the focal plane
of HRCam.  The probe is moved laterally on its translation stages, and
its  images  are  recorded  with  HRCam. The  position  of  the  image
centroids is  approximated by the linear function  of the coordinates,
relating  the detector pixels  to the  focal plane  coordinates.  Such
relation is  also determined for  the regular SAM imager  covering the
3$'$  field.   Image  orientation  on  the  sky  determined  from  the
astrometric solution of the  imager can thus calibrate the orientation
of the HRCam detector.

\subsection{EM CCD detectors}
\label{sec:det}

Table~\ref{tab:det} gives some characteristics of the Luca DL 658
(hereafter Luca) EM CCD used since 2007 and the iXon X3 888 (iXon-888)
camera used in 2017. Both cameras are manufactured by Andor.\footnote{www.andor.com}

\begin{table}[ht]
\center
\caption{Characteristics of the Luca and iXon-888 EM CCDs}
\label{tab:det}
\begin{tabular}{lcc}
\hline
Parameter & Luca DL 658 & iXon X3 888 \\
\hline
Format H$\times$V [pixels] & 658$\times$496  & 1024$\times$1024 \\
Pixel size [$\mu$m] & 10 & 13 \\
QE(540nm) & 0.50 & 0.96 \\
QE(790nm) & 0.25 & 0.82 \\
Response [el/ADU] & 1.7   &  10.1   \\
EM gain & 1--300 &   1--1000 \\
Readout noise [el] &  15  &  45  \\
CIC [el/pixel] & 0.07 & 0.02 \\
\hline
\end{tabular}
\end{table}

The Luca uses the Texas  Instruments line-transfer CCD.  The charge is
stored in light-protected areas near each pixel, to be transferred and
amplified  after   the  end  of   the  exposure.  The   line  transfer
architecture allows a very short  exposure time without any image blur
associated with  the charge transfer.   This feature turned out  to be
very useful at  SOAR, allowing us to mitigate  telescope vibrations by
exposure  times as  short as 2\,ms,  if the star  is sufficiently
bright.  As any front-illuminated CCD,  the Luca detector has a modest
quantum efficiency (QE) peaking  at 0.5. The CCD is thermoelectrically
cooled to $-20^\circ$C.

\begin{figure}[ht]
\epsscale{1.0}
\plotone{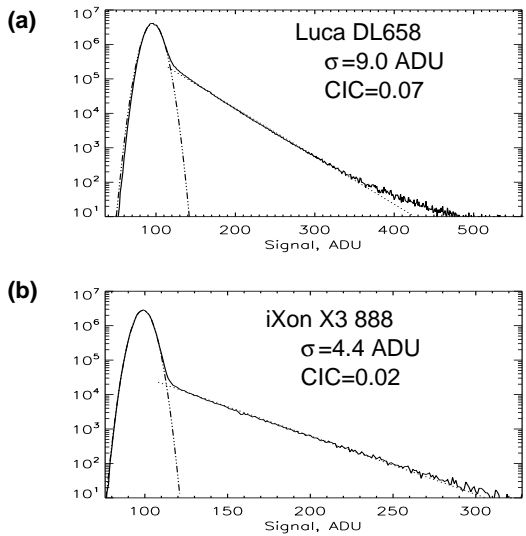}
\caption{Histograms of signal in the bias images and their
  models: (a) Luca, (b) iXon-888. The histogram is plotted by the full
  line, the two terms of the model (\ref{eq:hist}) by the dash-dot and
  dot lines. 
\label{fig:hist}
}
\end{figure}

Dark images  taken with  the EM gain  have a typical  appearance: most
pixels contain only noise, but  some isolated pixels are bright.  They
correspond to  the clock-induced-charge (CIC)  generated and amplified
in the EM register.  Noise parameters  of the EM CCD can be determined
from  the  distribution of  its  signal (Figure~\ref{fig:hist}).   The
signal histogram  can be modeled by  a sum of two  terms: the Gaussian
distribution  corresponding  to the  readout  noise  and the  decaying
exponent  that corresponds to  the amplitude  distribution of  the CIC
and photon events:
\begin{equation}
h(y) \approx  h_{01} \exp [ - \frac{(y - y_0)^2}{2 \sigma^2}] + 
h_{02} \exp [- \frac{y-y_0}{a}]. 
\label{eq:hist}
\end{equation}
Here $\sigma$  is the readout noise,  $a$ is the  typical amplitude of
the amplified single-photon events, both  in ADU.  The CIC rate is the
fraction of  pixels with  the signal level  above $5 \sigma$  from the
bias value, $y_0$ (in this case  about 100 ADU). The Luca camera has a
relatively  high CIC  rate  of 0.07.   Even  the 200$\times$200  image
fragment thus contains  about 3000 CIC events which  dominate over the
signal from faint stars and seriously limit the sensitivity of HRCam.

In 2014 July,  the Luca camera failed: it  simply lost any sensitivity
to light. The camera was sent for repair to the vendor and returned in
working  condition  in  early  2015.  However,  intermittent  failures
happened again in  2016 May, and the vendor  suggested that the camera
cannot be  repaired anymore.  In the  observing runs of  2014 and 2016
December, we installed  on HRCam the Luca-R cameras  loaned from other
programs.   They  also  use   front-illuminated  EM  CCDs  from  Texas
Instruments, but with smaller 7.4\,$\mu$m pixels (L2 of $F=75$\,mm was
then installed)  and with the frame-transfer  architecture; the format
is 1004$\times$1002  pixels. We found  that these cameras have  a poor
charge transfer efficiency (CTE)  in the vertical direction, resulting
in the  loss of resolution. The  blur depends on the  signal level: it
reaches 5-6 pixels for faint  stars, but becomes negligible for bright
ones. This signal-dependent  blur had to be accounted  for in the data
processing,  as  described  in  \citep{SOAR14}.   Obviously,  the  CTE
problem degraded the resolution and the measurement precision. The CIC
spikes in  the Luca-R CCD are  not blurred vertically by  the poor CTE
because they are  not produced in the CCD  pixels but rather generated
during the readout. 

In 2017,  we started to use  a much better iXon-888  camera, loaned to
SOAR by N.~Law  (UNC). This EM CCD is back-illuminated  and has a very
good QE  (Table~\ref{tab:det}).  Moreover, its detector  can be cooled
to $-80^\circ$C, resulting in  the negligibly small dark current.  The
optics and mechanics of HRCam  was adapted as described above. The new
detector was characterized by a series of tests.  Its EM gain actually
corresponds to  the gain setting  (unlike Luca). All  parameters match
the specifications except  the CIC rate, which was  found to be around
0.06 el/pixel, significantly exceeding  the 0.01 rate announced by the
vendor. However, the CIC rate could be reduced to 0.02 by reducing the
vertical  transfer  time to  3.3\,$\mu$s,  faster  than the  ``minimum
recommended''  time of  6.5\,$\mu$s.   With this  setting, the  charge
transfer was still  perfect, but with an even  faster clock the charge
transfer stopped working and the CCD produced no images.

The  iXon-888 camera  contains  a  frame transfer  EM  CCD, hence  the
minimum exposure time is restricted by the readout rate of 10\,MHz per
pixel.    For  the  normally   used  region   of  interest   (ROI)  of
200$\times$200 pixels (without binning),  the minimum exposure time is
24.4\,ms, and the fastest frame time is 27.9\,ms.  This exposure, used
mostly in  2017, makes  the results sensitive  to the 50  Hz telescope
vibration.   With the  2$\times$2 binning,  the same  ROI can  have an
exposure time  of 13.5\,ms  (i.e. two times  faster).  An  even faster
operation  is possible  in  the so-called  cropped-sensor mode,  where
vertical  stripes   of  selected  width  are  shifted   and  read  out
continuously.   In  the cropped-sensor  mode,  the 200$\times$200  ROI
without binning can be exposed for 6.7\,ms.  However, in this mode the
star must  be located at 100 pixels  from the left edge  of the field,
not  at the  center.   The cropped-sensor  operation was  successfully
tested on  the sky, but  not used routinely because  switching between
the  modes cannot  be  done  rapidly and  thus  affects the  observing
efficiency.

\subsection{Computers and Software}
\label{sec:SW}

The digitized video signal of  the Luca camera is acquired through the
USB interface.  As  the data-taking computer was located  far from the
instrument,   we  used   the  fiber-optics   signal   extender.   This
configuration  occasionally  had  connection  problems.   In  2015  we
replaced the standard data-taking computer by the compact Intel NUC PC
located  at the  telescope in  the electronics  rack, with  direct USB
connection to the camera.  This has improved the reliability. However,
this  PC had  no space  for the  PCI interface  board of  the iXon-888
camera that uses the  Cameralink communication protocol. Moreover, the
length of the Cameralink cable is  only 2\,m, forcing us to locate the
newly purchased fan-less PC near the HRCam. 

The HRCam  software was developed by R.~Cantarutti  using the Software
Development  Kit (SDK)  provided by  Andor, as  well as  LabView.  The
software allows selection of the  EM gain, exposure time, binning, and
the ROI.  Several settings of the ROI and binning (detector modes) are
defined   in  the   configuration  file   and  normally   used  during
observations.  Once the detector  parameters  are set, the  images are
acquired continuously  in the ``run  to abort'' mode and  displayed in
real time, for centering and focusing.  The desired
number of  sequential frames can be  written as an  image cube (16-bit
integer numbers) into the  FITS file.  Its header contains information
from  the  SOAR  Telescope  Control  System (TCS)  and  from  the  SAM
instrument, as well as the settings of the HRCam itself.  The software
has  a  convenient  graphical  user  interface  (GUI).   Moreover,  it
provides for the  display of the acquired cubes  using the DS9 utility
and an optional  calculation and display of the  power spectrum.  This
quick-look analysis capability is essential for evaluation of the data
quality.  When  a new binary or  triple system is  discovered, this is
usually  immediately   recognized,  allowing  the   observer  to  take
additional data for confirmation.

Table~\ref{tab:runs} gives the synopsis  of the observing runs and the
evolution  of the instrument  and observing  technique with  time. The
column AO indicates  the use of AO correction for  some targets in the
corresponding runs.


\begin{table}[ht]
\center
\caption{Summary of observing runs}
\label{tab:runs}
\begin{tabular}{c c c l}
\hline
Dates & Camera & AO & Notes \\
\hline
2007.81 -- 2007.82  & Luca  & No & First HRCam run \\
2008.53 -- 2008.55  & Luca & No &  Blanco run \\
2008.60 -- 2009.26   & Luca  & No & Without ADC \\
2009.66 -- 2011.07   & Luca  & Yes & SAM in NGS mode \\
2011.28 -- 2014.31  & Luca  & No &   \\
2014.77 -- 2014.86  & Luca-R & No &  Luca-R, poor CTE \\
2015.03 -- 2015.92  & Luca  & No & Start using SAA \\
2016.04 -- 2016.05  & Luca & Yes & Young stars \\
2016.13 -- 2016.14  &  Luca  & No &  \\
2016.38 -- 2016.40  & Luca  & Yes & Kepler targets \\
2016.94 -- 2016.97  & Luca-R & No &  Luca-R, poor CTE \\
2017.28 -- 2017.83  & iXon-888   & No & New camera \\
\hline
\end{tabular}
\end{table}

\section{Observations and data processing}
\label{sec:obs}

\subsection{Observing procedure and tools}
\label{sec:OT}

Accumulation of the standard cube  of 400 frames takes only 11\,s. The
observing  efficiency mostly  depends on  the time  used to  point the
telescope and to set the instrument parameters. As has always been the
case in  speckle interferometry, careful preparation  of the observing
program  and an efficient  strategy are  key ingredients  for reaching
high productivity.  Previous speckle  programs on 4 m telescopes could
observe up to 200 stars  per night; for example, 775 stars were measured
  in 4 nights at CTIO by \citet{McAlister1990}.

The software for planning  and executing HRCam observations is written
in IDL.  The observing program database contains essential information
on  all stars:  names, equatorial  coordinates, proper  motions (PMs),
magnitudes of  the components,  binary separation, and  short comments
indicating the reason for the  observation and the priority.  The date
of the last measure is also  stored and refreshed when new data become
available.   Tools exist  for adding  new  objects to  the program  by
retrieving information  from the  Washington Double Star  Catalog, WDS
\citep{WDS}, from  the {\it Hipparcos}  catalog, or from a  text file.
Objects to  be observed in the  forthcoming run are  selected from the
general database.  The program  of each run always contains additional
backup targets that can be observed under poor conditions.
 
Coordinates of   selected objects are computed for  the date of the
observation, accounting for the PM, and formatted into a list, grouped
by their position  on the sky. Originally, the  lists were loaded into
the TCS, and  the telescope operator was asked  to point the telescope
to  the next  target.  Starting  from  2014, this  procedure has  been
automated using the  new observing tool (OT). The  OT displays part of
the  sky around  the selected  target in  the  horizontal coordinates,
showing  the adjacent  targets. The  size of  the displayed  region is
selectable. The next target can be chosen by clicking in this display;
its  parameters  (and,  if  needed,  all  previous  measurements)  are
shown. By  pressing the button  in the OT  GUI, the observer  sends the
coordinates to the TCS, while the name of the target is entered in the
HRCam GUI. The  telescope slews to the new  target automatically if it
is within 15\degr ~from its previous position; otherwise, the slew must
be confirmed  by the telescope operator. The  OT substantially improves
the productivity and, at the same time, reduces the stress of both the
observer and  the telescope operator, as  well as the  number of human
errors. 

The  pointing of  SOAR is  good to  $\sim$5\arcsec ~rms.  However, the
field of view (FoV) of HRCam  is quite small, only 15\arcsec ~with the
1024$^2$ CCD.  For target acquisition,  the full camera field is used,
with a small EM gain and  an exposure time of 0.2\,s.  When the target
is centered, the  detector mode (ROI and binning),  the exposure time,
EM gain, and  filter are selected.  The zero-point  command is sent to
the TCS to refine the pointing,  so that the next nearby star is often
acquired in the same ROI without  looking at the full field. A special
command  in the SAM  control software  sets the  ADC according  to the
telescope coordinates and  flattens the DM.  The SAM  can also command
telescope  offsets  in  position  and  focus. The  focus  is  adjusted
visually by observing the star in real time.  A sequence of targets in
the same area of the sky can be observed rapidly without any action of
the telescope  operator. In  2017 June, 306  stars were  observed with
HRCam in one night.

Normally, two  data cubes, of 400  frames each, are  recorded for each
target  and  each  filter.   Given  the  small  time  needed  for  the
acquisition  of the  extra cube,  this  practice does  not affect  the
efficiency. Two  (or more) cubes  are processed independently  and the
final  results are  averaged, while  their mutual  agreement  gives an
estimate  of  the  internal  error.   Acquisition of  two  data  cubes
guarantees from ``glitches''  such as an occasional cosmic  ray in one
of the frames and makes the detection of new companions more secure.

The    standard    detector    mode    is   a    200$\times$200    ROI
(3\arcsec$\times$3\arcsec ~on the sky)  without binning and 400 frames
per data cube. Fainter stars  are sometimes observed with a 2$\times$2
binning  and   an  increased   exposure  time.   This   increases  the
sensitivity at the expense  of degraded resolution. Binary stars wider
than 1\farcs5  are observed with  a wider 400$\times$400 ROI  (with or
without binning) to avoid image truncation and aliasing. The wider ROI
is also  useful on the nights  with a strong wind,  when the telescope
shake  throws the  star outside  the 3\arcsec  ~field (we  do  not use
guiding). The vast majority of observations are made in the $I$ or $y$
bands, while other filters are used only occasionally. 

The image  delivered by  the telescope can  be sharpened using  the UV
laser and the SAM AO system.   As HRCam is mounted on SAM anyway, this
option comes  for ``free''.  In this  regime, we do  not acquire guide
stars with  SAM, while  the acquisition of  the laser and  closing the
laser loop are  fast (about a minute for large slews  or a few seconds
for  small slews). However,  the lists  of the  laser targets  must be
submitted in advance  to the Laser Clearing House  for approval of the
laser propagation. Laser-assisted speckle  runs were done for programs
with faint targets \citep{Kepler,BT17}.

\subsection{Data cube processing}
\label{sec:cube}

The data cubes are processed by the custom IDL software to compute the
power spectrum (PS) and auxiliary images. The PS is the square modulus
of the Fourier Transform (FT) of the intensity distribution in each
frame, averaged over all frames in the cube.  

The optimum  way to calculate the PS was found  by experimenting with
the real data.  To reduce the impact of noise  in the ``empty'' pixels
that do not contain star photons, the image is thresholded at $\sim$20
ADU above  the bias level. Signal  histograms in Figure~\ref{fig:hist}
show that such thresholding indeed cuts most of the readout noise. 

Images taken with  the Luca camera contain a  number of ``hot'' pixels
where the  thermal signal  is detectable even  at short  exposures. To
account for  this, two bias  image cubes with different  exposure time
(e.g.  20\,ms  and 100\,ms) are taken and  median-averaged. From these
data, the bias  image corresponding to the zero  exposure time and the
dark current (in ADU/s) in each pixel are computed, for the full frame
and the fixed EM gain used in the observations. Their combination with
the actually used  exposure time and ROI parameters  is the bias level
for each data frame, to be subtracted before the thresholding. If such
correction is not done, hot  pixels are clearly visible in the average
images of faint stars. We  do not apply flat-field corrections because
the CCD sensitivity variations of a few per cent are much smaller than
the speckle noise and their correction does not improve the quality of
the final data products. 

The dark  current of the  iXon-888 camera operated at  $-60^\circ$C is
negligible, so  there is  no need  to account for  hot pixels.  On the
other  hand,  the bias  level  in its  images  has  a vertical  (along
columns) structure that does not depend  on the EM gain but depends on
the ROI and  binning. To correct for this, bias  image cubes are taken
without EM gain for  each ROI/binning combination. The median-averaged
bias signal is subtracted from each column of the data frames.

The PS  of speckle images of  a point source, $P_{0}({\bf  f})$, has a
characteristic two-component structure
\begin{equation}
P_{0}({\bf f}) \approx | T_{\rm SE} ({\bf f}) |^2 + 0.435 (D/r_0)^{-2}  T_0 ({\bf  f}),
\label{eq:ps}
\end{equation}
where $  T_{\rm SE} ({\bf  f}) $ is the  seeing-limited short-exposure
transfer function (FT of the  average re-centered image), $r_0$ is the
Fried parameter, $D$  is the telescope diameter, and  $ T_0 ({\bf f})$
is  the transfer  function of  an ideal  diffraction-limited telescope
\citep[see  e.g.][]{Christou1985}.   The  second  term  describes  the
high-frequency  part of  the  PS,  i.e. the  speckles.   In the  image
auto-correlation  function,  ACF  (FT   of  the  PS),  the  first  term
corresponds  to  the broad  seeing-limited  component, usually  called
``seeing pedestal'', while the  high-frequency term is responsible for
the  narrow diffraction-limited  peak  at the  coordinate origin,  the
``speckle peak''.

The  PS   of  a  binary  or  multiple   star  contains  characteristic
fringes.  The  ACF  has   corresponding  secondary  peaks  at  the
separation   $\rho$  and   position   angle  (PA)   $\theta$  of   the
companion.  However,  the  PS  contains  most energy  at  low  spatial
frequencies  that produce the  seeing pedestal  in the  ACF.  The
HRCam pipeline removes the seeing  pedestal by computing the ACFs from
the spatially  filtered PS \citep{TMH10}.  The ACFs  play an important
role in the  data processing. Wide binary companions  are more readily
detected in the ACF than in the PS, while ``fringes'' produced by very
close companions are more obvious in the PS.

The ACF  is computed from the  PS, i.e. the average  square modulus of
the image  FT. This non-linear  transformation extends the  support of
the ACF  to twice the FoV  size $\Omega$.  A  correct ACF calculation,
not implemented  in the standard  pipeline, should use  the re-sampled
PS.  This  matters only  for wide binaries  with separation  $\rho \ge
\Omega/2$.  In  such cases,  the ACF  computed by a  simple FT  of the
(filtered) PS may have the companion's peaks in the wrong place due to
aliasing. For this reason, wider  FoV is used for observations of wide
pairs; this also reduces the image truncation.

Apart from the PS, the  pipeline computes the  average image and
the re-centered average image.  A simple centroid algorithm works well
in most cases. However, for  faint stars, when the object flux becomes
comparable  to the  CIC rate,  the spurious  events seriously  bias the
centroid. Centroiding of  the smoothed and thresholded version of
each frame  works much better  \citep[see][]{Kepler}. Centroiding also
helps to reject frames where the star  is too close to the edge of the
field (e.g. because of the telescope wind shake).

Since 2015, the speckle pipeline also computes the shift-and-add (SAA)
image  centered  on  the  brightest  pixel in  each  frame.   This  is
analogous  to  the  ``lucky  imaging'',  except  that  no  frames  are
rejected.  A weight proportional to  the signal in the brightest pixel
is applied to  each frame. These SAA images  often contain the central
diffraction-limited peak and the  secondary peaks corresponding to the
binary  companion.   Selection  of  the brightest  peak  resolves  the
180\degr   ~PA   ambiguity    inherent   to   the   standard   speckle
processing. The SAA images are  much noisier than the standard speckle
ACFs because they do not  use the information optimally \citep[see the
  comparison  between ACF  and SAA  in][]{SAM09}.  The  SAA  images of
faint stars, centered on random  photon spikes rather than on the real
speckles, are useless, while binary companions are still detectable in
the PS and ACF of these stars.

\begin{figure}[ht]
\epsscale{1.1}
\plotone{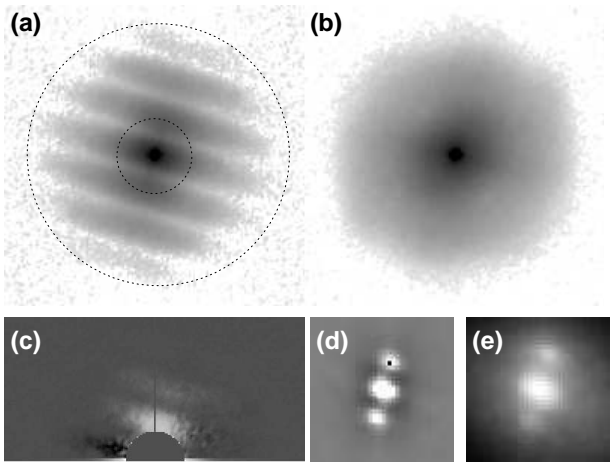}
\caption{Example   of   data   processing.    The  binary   star   WDS
  J01376$-$0924 (KUI~7) was observed in the $y$ filter on 2017.83 with
  the  iXon-888 camera.   Binary parameters:  $\rho  = 0\farcs1229$,
  $\theta = 343\fdg4$,  $\Delta m = 1.22$ mag. Panel  (a) shows the PS
  in  negative logarithmic  stretch, where the dotted circles mark the
  fitting zone between 0.2 and 0.8 $f_c$, (b)  is the  reference PS  of the
  unresolved star, (c) is the residual  to the model (upper half
  of the frequency plane  in linear stretch), (d) is the central fragment
  of the ACF  (the black point marks the companion's  peak), and (e) is
  the fragment of the SAA image.
\label{fig:images}
}
\end{figure}

To  summarize,  the  pipeline   produces  from  each  data  cube  four
two-dimensional    images:     PS,    ACF,    centered,     and    SAA
(Figure~\ref{fig:images}).   The  headers   of  those  images  inherit
information  from  the  original  image cube.  This  information  also
populates  the database that  holds results  of the  observations (see
\S~\ref{sec:data}).

\subsection{Fitting binary and triple stars}
\label{sec:bin}

Processing of  HRCam data  is described by  \citet{TMH10}. Here  it is
briefly  recalled  with  an  emphasis  on  the  caveats.   First,  the
photon-noise bias  in the PS is  determined by averaging  it over the area
beyond the  cutoff spatial frequency  $f_c = D/\lambda$  ($\lambda$ is
the wavelength, $D$ is the  telescope diameter). It is subtracted from
the PS and accounted for while computing the noise.

The  PS of  a multiple  star with  $M$ components  (point  sources) is
approximated by the model $P_{mod}({\bf f})$
\begin{equation}
P_{mod}({\bf f}) = P_0({\bf f}) \; | \sum_{i=1}^M a_i \exp ( 2 \pi {\bf f}
{\bf x}_i) |^2 ,
\label{eq:bin}
\end{equation}
where  ${\bf f}$  is  the  spatial frequency,  $P_0({\bf  f})$ is  the
reference  PS of  a  single star,  $a_i$  are the  intensities of  the
components and ${\bf x}_i$ --  their coordinates. The PS is normalized
to $P(0) =1$, which translates to  $\sum_i a_1 = 1$.  Moreover, the PS
is invariant to the translation  of the source.  Therefore, a multiple
system with $M$ components  has $3(M-1)$ free parameters. Although the
formula  (\ref{eq:bin})  looks  simple,  the square  modulus  contains
cross-terms between  all components,  so the analytical  expression of
the  derivatives  of  the   model  over  parameters,  needed  for  the
model-fitting, becomes complicated  with increasing $M$. Model-fitting
is currently implemented only  for binary and triple stars, delivering
3 and 6 parameters (PA, separation, and $\Delta m$), respectively, and
their  errors.   For  stars  with 4  resolved  components,  additional
positions and  magnitude differences can be measured  crudely from the
peaks in the ACF.

For binaries with  large angular separations $  \rho \gg \lambda/D$,
the PS contains multiple fringes.   As a result, the binary's parameters
are decoupled from the shape of the reference PS and the result of the
fitting is very robust. In such cases, the rotationally-averaged PS of
the object itself makes a good  reference $ P_0({\bf f}) $ because the
fringes  are  effectively removed  by  the  averaging.  For  close
binaries with  separations on the  order of $\lambda/D$,  the analytic
model  of $P_0$  is used  for  fitting. This  model,  introduced in
\citep{TMH10}, is

\begin{equation}
P_{0, syn}({\bf f}) =  T_0(|{\bf f}|) 10^{-[p_0 + p_1 (|{\bf f}|/f_c)]} ,
\label{eq:P0syn}
\end{equation}
where $ T_0(f)$ is the  transfer function of the ideal telescope. This
model  is valid  only at  high spatial  frequencies $  \lambda/r_0 \ll
|{\bf f} | < f_c$.  It  is based on the  theoretical expression
for  the  speckle  transfer  function in  the  high-frequency  domain
(the second term of eq.~\ref{eq:ps}).  The  two  parameters of  the
synthetic  PS, $p_0$  and  $p_1$,  are determined  by  fitting  the
rotationally-averaged PS. When the  speckle structure is a perfect match
to the theoretical model, $p_1 = 0$ and
\begin{equation}
(D/r_0)^2  = 0.435\; 10^{-p_0} .
\label{eq:Dr0}
\end{equation}
In  reality, the  finite spectral  bandwidth and  the  finite exposure
time,  as well  as vibrations,  reduce  the speckle  contrast at  high
spatial frequencies, leading to the  faster PS decay, hence $p_1 > 0$.
However, the parameter  $p_0$ is still a valid  measure of the average
number of speckles, and the expression (\ref{eq:Dr0}) holds, allowing
us to compute $r_0$ and hence the seeing.

As the PS is symmetric, the binary or triple models are fitted only to
the upper half of the  frequency plane and in the restricted frequency
range $ 0.2 f_c < | {\bf f} | < f_{\rm max}$. By default, $f_{\rm max}
= 0.8 f_c$,  but it is reduced for noisy  data.  Weights of individual
pixels in the frequency plane  are inversely proportional to the noise
variance in these pixels, calculated analytically.  The quality of the
fit is determined by the  usual $\chi^2/N$ parameter computed from the
residuals and the noise. A  perfect model results in $\chi^2/N \approx
1$.  In practice, this happens for faint stars, while large $\chi^2/N$
are  found for  bright stars,  where  the residuals  are dominated  by
un-modeled details of the PS.

Since 2015, we can use PS of other stars (both single and binary) as a
reference  \citep{SOAR15}.   In the  latter  case,  the fitted  binary
parameters serve to deconvolve  the reference PS from the fringes,
at the expense  of the increased noise. Only  binaries with fringes of
low  or  moderate  contrast ($\Delta  m  <  1$  mag) are  suitable  as
reference. Naturally, the object and the reference must be observed in
the same filter,  with the same detector format, close  in time and in
the sky.  Use of the real reference is particularly helpful in fitting
difficult cases,  such as binaries  with small separation and  a large
$\Delta m$.   As shown in Figure~\ref{fig:images}, even  with the real
reference  the  residuals  rarely  resemble  a white  noise  owing  to
systematic  differences  between  the  data  and the  model;  in  this
particular case, $\chi^2/N =13.3$.

In  2008 and 2009,  when HRCam  worked without  AD correction,  the PS
model included speckle elongation caused by the AD. The elongation was
computed from the  known central wavelength and the  bandwidth of each
filter,  knowing the telescope  pointing. In  2014, the  vertical blur
caused  by  the  poor  CTE  had   to  be  included  in  the  PS  model
\citep{SOAR14}.

\subsection{Detection limit, resolution, and sensitivity}
\label{sec:dmlim}

\begin{figure}[ht]
\epsscale{1.0}
\plotone{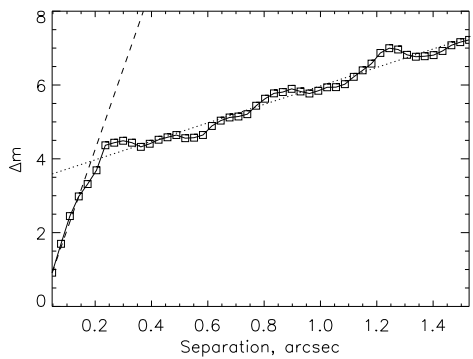}
\caption{Detection limit   for an  unresolved star
  shown  in Figure~\ref{fig:images}  (b);  $y$ filter,  200$\times$200
  ROI.The dashed  and dotted lines are linear  approximations at small
  and large separations, respectively.
\label{fig:det}
}
\end{figure}

The detection  limits are estimated  from the fluctuations of  the ACF
computed  in annular  zones of  increasing radii.   The  rms amplitude
$\sigma$ is converted into the maximum detectable magnitude difference
$\Delta  m$  by  assuming  that  peaks  larger  than  $5  \sigma$  are
detectable. This assumption has  been verified on simulated companions
\citep{TMH10}.  As shown  in Figure~\ref{fig:det}, the detection limit
$  \Delta m  (\rho)$ increases  rapidly at  small separations  $\rho <
0\farcs2$  and  then continues  to  improve  more gradually,  reaching
$\sim$6  mag for  good-quality data.   The dynamic  range of  HRCam is
comparable to other speckle instruments.  \citet{Horch2012} can detect
companions with $\Delta m \sim  5.5$ mag at 0\farcs2 separation (their
Fig.~5), on a larger telescope and with a larger number of accumulated
frames.    To  give   an  example,   the  subsystem   EHR~9~Ba,Bb  (WDS
J06454$-$3148) is separated by 1\farcs4 from the main component A, and
its components are  $\sim$6.5 mag fainter than A in  the $I$ band.  It
was measured  with HRCam 10 times,  leading to the  calculation of the
orbit  with  a  period  of  7 years  \citep{Tok2017}.   Note  that  at
separations $\rho  > 1$\arcsec ~the speckle signal  (and the detection
limit)  is  reduced   by  the  anisoplanatism;  this  seeing-dependent
reduction is  not accounted  for in the  computed curves like  that in
Figure~\ref{fig:det}.  For binary stars, the curves $ \Delta m (\rho)$
show sharp dips at the companion's separation.

The curves $ \Delta m (\rho)$  at small and large separations are well
approximated by  two linear functions. The parameters  of these linear
fits are  stored in {\tt  obsres} (see \S~\ref{sec:data}) and  used to
compute  the detection  limits at  other separations.   For unresolved
stars, the published  data tables give these limits  at separations of
0\farcs15 and 1\arcsec.

The nominal  angular resolution  of speckle interferometry  equals the
diffraction limit $\lambda/D$, i.e.  27\,mas at 540\,nm and 40\,mas at
800\,nm. At this  separation, the maximum of the  second fringe in the
PS  of  a binary  star  coincides  with  the cutoff  frequency  $f_c$.
However, when the data are of good quality (with the speckle signal at
$0.8 f_c$ exceeding the noise), even closer binaries can be measured by
fitting  the PS  model.  Separations  as  small as  12\,mas have  been
measured at  540\,nm.  On the  other hand, for  faint stars the  PS is
lost in the noise well before  the $f_c$ is reached, and the effective
resolution  is  substantially worse  than  $\lambda/D$.  The  pipeline
accounts for  this by reducing  the $f_{\rm max}$, and  the resolution
limits in the tables of unresolved stars are increased proportionally.
However,  the  estimated  resolution  limits remain  approximate  and,
possibly, optimistic.

The sensitivity  (limiting magnitude)  of speckle interferometry  is a
strong function of  the seeing blur $\beta$ (or,  equivalently, of the
Fried  parameter $r_0  = 0.98  \lambda/\beta$) because  the  number of
speckles  is proportional  to $(D/r_0)^2$,  i.e. to  $\beta^2$.  Faint
stars  are usually  observed in  the $I$  band because  of  its larger
bandwidth (hence larger flux) and  the larger $r_0$.  In practice, the
magnitude limit reaches $I =  12$ mag under good seeing. Still fainter
stars can be  observed with a longer exposure time,  at the expense of
the spatial  resolution.  Considering that the seeing  is variable and
that the signal  can be further degraded by  vibrations, the magnitude
limit cannot  be guaranteed in the forthcoming  observing runs.  Under
very  poor   conditions  (poor  seeing   and/or  transparent  clouds),
good-quality measurements of bright stars are still possible.

The new  iXon-888 camera  exceeds the sensitivity  of Luca in  the $I$
band by at  least a factor of two, owing to  its larger QE. Additional
gain is  provided by its lower CIC  (Table~\ref{tab:det}). This camera
is used since  2017 April. During this period,  only one half-night of
good seeing was experienced. On that night, a binary star with $I =14$
mag was resolved, demonstrating the increased sensitivity of iXon-888.

The image  size $\beta$ can  be improved by  the SAM AO  system.  This
observing mode  was used twice  for programs with  predominantly faint
targets  \citep{Kepler,BT17}.  With  a  longer exposure  time and  the
2$\times$2 binning, not  much is left of the  speckle signal. However,
the re-centered images  are still quite sharp and  allow discovery and
measurement of binaries down to $\sim$0\farcs1.

\subsection{Artifacts}
\label{sec:vibr}


\begin{figure}[ht]
\epsscale{1.0}
\plotone{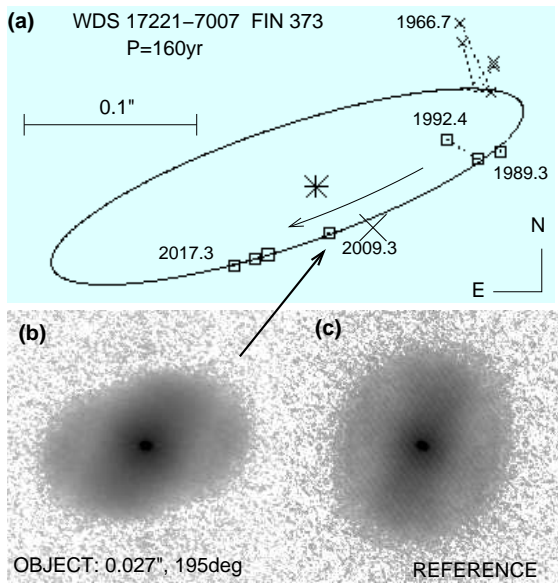}
\caption{The orbit  of the  close  binary WDS  J17221$-$7007 (FIN~373)  is
  shown in (a).  Squares  denote speckle measurements, crosses are the
  historic visual measurements by W.~Finsen. The published measurement
  made  with HRCam  in  2009.3  (large cross)  is  wrong because  the
  ``fringe''  in  the  PS  (b)  caused  by  vibration  was  originally
  interpreted as the binary-star  signature. Use of the real reference
  (c)  produces the  correct  measurement with  $\rho  = 27$\,mas  and
  $\theta = 195\degr$.
\label{fig:FIN373}
}
\end{figure}

It is well known that  small vibrations of the telescope optical axis,
irrelevant for  seeing-limited observations,  can be very  detrimental to
speckle  interferometry. Unfortunately,  the SOAR  telescope  is often
affected by  vibrations with  a frequency of  50\,Hz and  an amplitude
reaching 30\,mas. They  are present in the signals  of the guiders and
in the  AO data  recorded by  SAM. The optical  axis oscillates  on an
elliptical trajectory with variable eccentricity, from nearly circular
to nearly  linear. The amplitude  of these oscillation is  variable in
time and depends on the telescope pointing (larger at low elevations).
See \S~3  of \citep{SAM09} for  more information on  these vibrations.
Our current  understanding is that  they are excited by  vibrations of
the soil  with the 50\,Hz  frequency produced by  electrical equipment
such  as transformers.  The  50 Hz  signal is  indeed detected  by the
accelerometers installed at  the telescope pier and at  the top end of
the telescope itself.   However, the amplitude of
these  mechanical vibrations  is an  order of  magnitude too  small to
explain the  oscillations of the  optical axis, and their  waveform is
not elliptical.  Plausibly,  the servo-controlled fast tip-tilt mirror
of SOAR amplifies the 50  Hz mechanical perturbation under some, still
unidentified, conditions.

The  elliptical blur  of speckles  caused by  the vibrations  leaves a
characteristic signature  in the PS;  for quasi-linear motion,  the PS
acquires a  fringe-like structure  and can mimic  a binary  star.  One
such  difficult   case  is  illustrated   in  Figure~\ref{fig:FIN373}.
Fortunately, another  star with a similar vibration  distortion can be
used  as a reference  and allows  to measure  the parameters  of close
pairs, which often happen to be critical for the orbit calculation.

The  vibrations   reduce  the  high-frequency  power   and  hence  the
sensitivity  and/or resolution.  Bright  stars were  normally observed
with the  exposure time  of 5\,ms or  shorter, possible with  the Luca
camera. This  strategy recovers the  resolution at the expense  of the
sensitivity. With  the new iXon-888 camera, the  typical exposure time
is 24\,ms,  so we  can only  count on the  intermittent nature  of the
vibrations. Sometimes they vanish completely and the PS extends almost
to the cutoff  frequency, as in Figure~\ref{fig:images}.  Observations
of the same  star with a different exposure time  is a good diagnostic
of the vibrations.

\begin{figure}[ht]
\epsscale{1.0}
\plotone{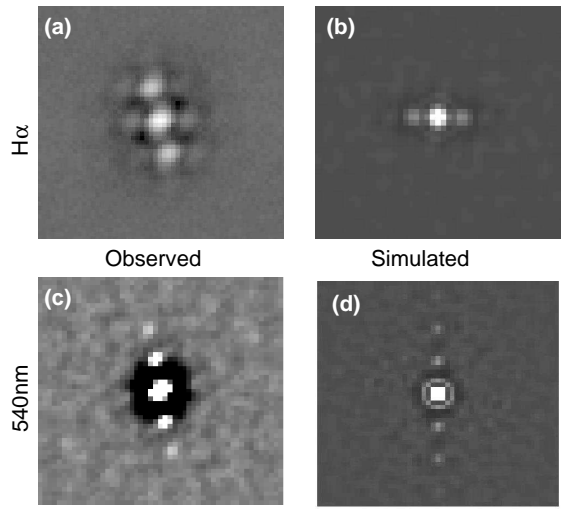}
\caption{Optical  ghosts  and  their   simulation.   The  ACF  of  the
  0\farcs12 binary with  OGs observed on 2008 August  in the H$\alpha$
  band  is shown  in  (a); each  peak  is surrounded  by two  spurious
  maxima. The OGs are  simulated  in (b) by combining seeing with the sine
  phase wave-front distortion of  0.4\,$\mu$m amplitude and 2 m period.
  The  OGs with two maxima  recorded in 2016  February at 540\,nm
  are shown  in (c) for a single  star.  The simulated ACF  in (d) is
  produced by  a combination of seeing  with the clipped  sine wave of
  0.2\,$\mu$m amplitude and 1 m period.
\label{fig:ghosts}
}
\end{figure}

Yet another  phenomenon that  can mimic a  binary star  is encountered
sometimes under conditions of slow wind.  Each peak in the ACF is then
surrounded  by two  spurious faint  peaks,  often located  at $\sim  2
\lambda/D$ separation,  near the first diffraction  ring.  Unlike real
binary companions, the separation  of these {\it optical ghosts} (OGs)
is  proportional  to  the  wavelength, hinting  on  their  diffraction
nature.  Indeed,  the OGs  can be reproduced  in simulation  if random
atmospheric wave-fronts caused by the seeing are combined with a fixed
periodic phase screen.  Examples of the real and  simulated OGs in the
top row of Figure~\ref{fig:ghosts} are taken from the data obtained in
2008 \citep{TMH10}.  In 2016 February, the OGs with double diffraction
spikes were  seen, appearing and disappearing  intermittently during a
period of an  hour \citep{SOAR16}.  Those OGs correspond  to the phase
perturbations with a spatial period of $\sim$1\,m and a non-sinusoidal
shape.  The OGs are most likely produced in the air near the telescope
when the wind  speed is slow.  Their dependence  on the wavelength and
similar appearance  in different stars  helps to distinguish  OGs from
real binary companions. Otherwise, the  OGs can be mistaken for binary
companions with $\Delta  m \sim 3$ mag and separations  from 60 to 120
mas.

Apart from vibrations and OGs, the  shape of the PS is affected by the
residual optical  aberrations, especially under good  seeing and/or at
longer  wavelengths. The observed  PS always  has some  structure that
lacks  axial  symmetry.   This  structure  depends  on  many  variable
factors,  such  as the telescope focus.   Two objects observed in  a short
succession in the  same filter often have similar  PS structure. Using
one of the  stars as a reference for another helps  to account for the
PS asymmetry  and reduces its  influence on the  measurement accuracy.
Such structure is notable  in the PSs shown in Figure~\ref{fig:images}
(a) and (b).

\section{Results}
\label{sec:res}

Presently (2017 November), the  total number of observations made with
HRCam is  11903 (Table~\ref{tab:runs}). This includes  the 6-night run
at the Blanco  telescope in 2008, when HRCam was  used as a substitute
for  the USNO  speckle camera;  all  other observations  were made  at
SOAR. The total number of observed  objects is 4366; 2819 of those are
resolved  pairs,  for which  8985  measurements  were  made.  The  WDS
contains  approximately 250  close binaries  or  subsystems discovered
with  HRCam.  A  total  of  25 refereed  publications  use  the  HRCam
data.\footnote{         See         the        bibliography         at
  \\ \url{http://www.ctio.noao.edu/\~{}atokovin/speckle/} }

\subsection{Data management}
\label{sec:data}

\begin{figure}[ht]
\epsscale{1.0}
\plotone{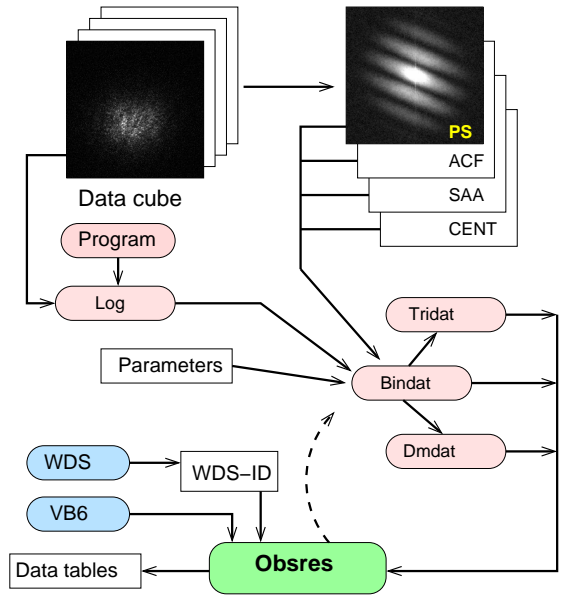}
\caption{Data flow in the speckle  pipeline. Colored ovals are  IDL
  structures, unfilled rectangles are  text files.
\label{fig:data}
}
\end{figure}

Reduction of  the data  cubes is  just the first  step in  the speckle
pipeline.  Managing  a large  number of observations  requires special
tools to do  this efficiently.  Figure~\ref{fig:data} illustrates the data
flow in the speckle pipeline, from raw image cubes to the final tables
of calibrated  measurements.  This process is explained  below. As the
pipeline is implemented in IDL, the results are stored and manipulated as
IDL  structures,  depicted  by the colored  ovals  in  the  Figure.   The
alternative, more traditional ways  of using FITS headers, text files,
or spreadsheets to store the processing results are less convenient.

Relevant information from the FITS  headers is stored in the {\tt log}
structure, one element  per data cube. It also  holds the intermediate
results (e.g. the  parameters $p_0$ and $p_1$ and  the total flux from
the object). The  {\tt log} is compared with  the observing program to
identify wrongly typed  object names and to add  a tag corresponding to
the particular program. It is checked for missing TCS information such
as date and coordinates. In the rare cases of such errors, the missing
information can  usually be  copied from the  second data cube  of the
same object. 

The {\it parameter  file}  plays a  key  role  in  the processing  of  each
observing  run. It  specifies  the directories  where  the images  are
stored,  the  calibration   parameters  (translation  from  the  pixel
coordinates to the sky coordinates),  and the files that hold the data
structures. Each observing run has its unique parameter file.  

The {\tt  bindat} structure, produced from  the {\tt log},  is used to
organize the binary-star processing. An  IDL GUI serves to display the
ACF, identify the  binary companion by clicking on it,  and to fit the
binary parameters.  It allows  to display other associated images (PS,
SAA, or centered)  by a simple click and to  navigate between all data
of the observing run. Unresolved stars are marked by setting the total
number of  components $M$ to one,  triple stars have  $M=3$.  The {\tt
  bindat} also  stores information on the  detection limits.  Although
the binary-star  fitting is  interactive, it can  be done  rapidly and
efficiently using the IDL GUI.

Stars  marked as  triple are  processed by  another GUI  program which
saves  the   results  in   the  associated  structure   {\tt  tridat}.
Alternative  photometry  of classically  resolved  binaries that  uses
centered  images  is  done  automatically, based  on  the  information
provided in  {\tt bindat}; its results  are stored in  the {\tt dmdat}
structure. 

The   correspondence  between   data   cubes  and   measures  is   not
straightforward.   On one hand,  two or  more data  cubes of  the same
binary star  are averaged  and produce only  one measurement.   On the
other hand, observation of a  triple star produces two measurements of
its subsystems  that must be  stored with different  names.  Published
measurements of binary stars should be provided with their identifiers
in  the WDS catalog  \citep{WDS} and  the standard  names (``discovery
codes''), if those exist.  So, the results of the processing contained
in {\tt bindat}, {\tt tridat}, and {\tt dmdat} are combined, averaged,
and  stored in  the  final  data structure,  {\tt  obsres}, where  one
element corresponds  to one observation  of a particular  subsystem in
one filter. 

After  the averaging,  each  observation has  its  internal WDS  code,
generated  from the coordinates,  and the  object name  inherited from
{\tt  bindat}.    Triple  stars  have  additional   internal  tags  to
distinguish between their two subsystems.  The WDS catalog, previously
transformed into an IDL structure,  is searched by coordinates to find
entries corresponding  to the observed  pair. If the WDS  contains several
binaries with the same code, the one with the best-matching separation
is  selected, while the  rest are  listed as  alternative suggestions.
The result of  this automatic identification is the  {\tt WDS-ID} text
file  that translates  internal  object names  into  the official  WDS
names.  For objects not found in the WDS, the internal names are kept.
This  ``dictionary''  needs  only   minor  manual  edits  because  the
automatic match  succeeds in  most cases.  The  dictionary is  used in
creating  the {\tt  obsres} structure,  where the  objects  have their
official names  and WDS  codes.  The file-name  of the  first averaged
data  cube is  kept  in the  {\tt  obsres} as  well, associating  each
measure with the images. The text files of the data
tables are generated (exported) from the {\tt obsres} structure.

The  {\tt  obsres} structure  is  the  final  product of  the  speckle
pipeline. Its elements are identified with the {\it Hipparcos} catalog
to create alternative object names. The Sixth Catalog of Visual Binary
Star Orbits, VB6  \citep{VB6}, is searched for orbits  of the observed
binaries to compute  the ephemeris positions and to  compare them with
the measures.   There is a GUI  program for browsing  and editing {\tt
  obsres}.  It allows to look  back at the corresponding images.  This
is done by associating the  date of the observation with the parameter
file of the  corresponding observing run; then the  binary-star GUI is
called with these parameters and the file-name of the data cube.  This
capability helps  to examine  questionable measures and  to re-process
them, if necessary, thus updating the {\tt obsres}. 

The last  step in the speckle  data reduction is the  manual check for
errors  and inconsistencies that  almost always  happen in  large data
sets. For example, if, after pointing the telescope to a new target,
the observer forgot to change the object name, the observations of two
binaries will be averaged together, producing a wrong measure with a
large internal error.  This situation can be corrected during creation
of the  {\tt log} structure,  during the binary-star  processing, and,
finally, by accessing and re-processing the data from the {\tt obsres}
GUI. If  several orbits for a given  binary are found in  the VB6, the
choice  can  be made  manually.  If  the star  is  not  found in  {\it
  Hipparcos}, its alternative name is entered manually, too.

The results  of all observing runs  can be joined  (glued) together in
the  common {\tt obsres}  structure.  It  can be  consulted by  the OT
during  the   observations,  used   to  extract  measures   for  orbit
calculation, or  to compare the latest observations  with the previous
ones.  As mentioned above, at present it contains 11903 entries.

\subsection{Calibration}
\label{sec:cal}

Owing to the small FoV  of speckle cameras, calibration of their pixel
scale and orientation  has always been difficult.  For  the HRCam, the
following methods were tried:

\begin{itemize}
\item
Calibration against visual orbits. The  accuracy of most orbits in the
VB6 catalog is inferior to  the accuracy of the HRCam measurements, so
this  method,  used  only  for   the  very  first  observing  run,  is
questionable.   It  is still  adopted  by  some  speckle  programs,
however. 

\item
Calibration using interference fringes is a standard technique, in use
since the 1980s.  In the case of HRCam,  installation of a double-slit
mask in the telescope beam  is not practical.  In 2009, the instrument
was  calibrated  by a  two-beam  laser  interferometer  with a  0.5  m
baseline attached to the telescope spider \citep{SAM09}.

\item
Astrometric calibration of the SAM imager and its translation to the
HRCam was used several times (see \S~\ref{sec:opt}). 

\item
Internal calibration using several wide binaries with a well-modeled
slow motion is the preferred method described below. It was introduced in
2014 \citep{SOAR14} and is used since that time. It can be applied
retro-actively to all HRCam measurements. 

\end{itemize}

\begin{figure}[ht]
\epsscale{1.0}
\plotone{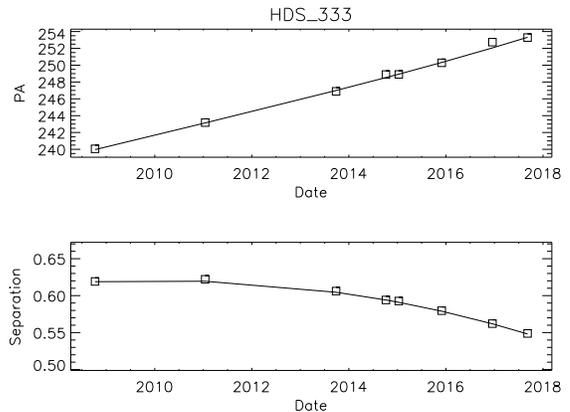}
\caption{Observations   of   the   calibrator binary WDS   J02332$-$5156
  (HDS~333)  in  PA  (top)  and  separation (bottom)  are  plotted  as
  squares, their models as  full lines. Eight measurements have the
  rms residuals  of 2.8\,mas  in tangential direction  and 0.9  mas in
  separation.
\label{fig:calib}
}
\end{figure}

\begin{figure}[ht]
\epsscale{1.0}
\plotone{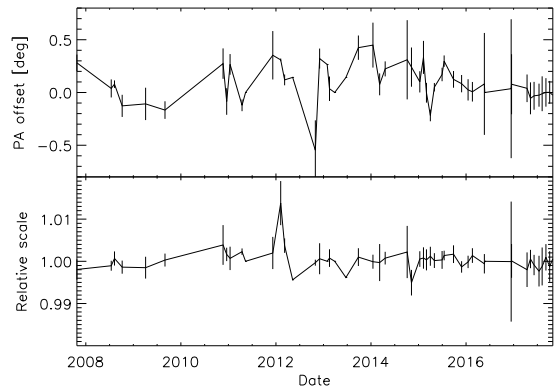}
\caption{Offsets in  PA $\Delta \theta$ and average  scale factors $s$
  for the 51 observing runs  made with HRCam since 2007, as determined
  from the calibrator binaries.  Vertical bars show the rms scatter of
  calibrators in each run.
\label{fig:plotruns}
}
\end{figure}

In 2017, the initial set of  calibrators was extended to 65 pairs with
separations from 0\farcs5 to 3\arcsec,  each observed at SOAR at least
3 times (on  average 8.6 times per binary).   Their motion was modeled
either by linear functions of time or by orbits, specially adjusted to
fit  the  SOAR   data.   One  such  pair,  HDS~333,   is  featured  in
Figure~\ref{fig:calib}. For each run,  the average correction in angle
$\Delta  \theta  =  \langle  \theta_{\rm  obs}  -  \theta_{\rm  model}
\rangle$ and the  average scale factor $ s =  \langle \rho_{\rm obs} /
\rho_{\rm  model}  \rangle$  are  determined.   After  applying  these
corrections, the models  of the binary motion are  refined.  After two
iterations  the process  has  converged.  The  rms  deviations of  the
corrected  measures from  the models  range from  1 to  3 mas  in most
cases.

Figure~\ref{fig:plotruns}  shows  the  calibration parameters  of  all
51 observing runs and  their rms scatter.  Typical speckle  runs have the
rms scatter  of the calibrators from  0\fdg1 to 0\fdg2 in  PA and from
0.002 to 0.004 in scale.  The  runs of 2016 May and 2016 December have
a larger than usual PA scatter of 0\fdg5 and 0\fdg3, respectively. The
most deviant point in the upper plot of Figure~\ref{fig:calib} is that
of  2016.96,  contributing  to   the  larger  rms  in  the  tangential
direction. Apparently, the control of  the Nasmyth rotator of the SOAR
telescope had some problems in  those two runs, degrading the accuracy
of the PA setting.

It is  desirable to  observe the same  calibrators with  other speckle
instruments in order to link their results with those of HRCam. Future
more accurate  measurements (e.g.  with  long-baseline interferometers
or from space)  may improve the calibration of  already published data
by using common binaries with well-studied motion.

\subsection{Orbit calculation}
\label{sec:orb}

The large  set of measurements  of close visual binary  stars obtained
with HRCam is used for calculation of their orbits and for improvement
of  already known orbits.   The VB6  catalog currently  contains more
than 400  orbits based  on the  HRCam data, amounting  to 15\%  of all
entries.        These        orbits       are       published       in
\citep{Gomez2016,Hrt2012,Mdz17,Tok2012,TGM14,planetary,SOAR14,Tok2016a,Tok2016b,Tok2016c,TL17,Tok2017}. Figure~\ref{fig:FIN373}
illustrates the orbit of the close binary FIN~373 based on four HRCam
measures  and two  previous speckle  measures.  The  prior orbit of  this pair
published in 2013 misinterpreted even  the sense of the orbital motion (in
fact it  is retrograde, i.e.  clock-wise).  \citet{Tok2016c} corrected
this aspect,  but computed an  orbital period $P=56.9$\,yr,  while the
longer period $P= 160$\,yr fits the data better.

In  the  least-squares  fitting  of  the  orbital elements,  the
weights  should  be  inversely  proportional  to  the  square  of  the
measurement errors.  Good visual  micrometer measures have an accuracy
reaching  20\,mas.  The  HRCam  data have  typical  errors of  2\,mas,
calling for the relative weight  of 100. Many visual measurements have
much larger errors, e.g. 0\farcs2, and their weights should be reduced
accordingly to $10^{-4}$.   The system of weights adopted  by the USNO
team \citep{VB6}  is much more  uniform.  As a result,  old inaccurate
data ``drag'' the orbit away from the best solution, and its residuals
to the modern measures are larger than their errors.  This is the case
for many orbits published in \citep{SOAR14}.

\begin{figure}[ht]
\epsscale{1.0}
\plotone{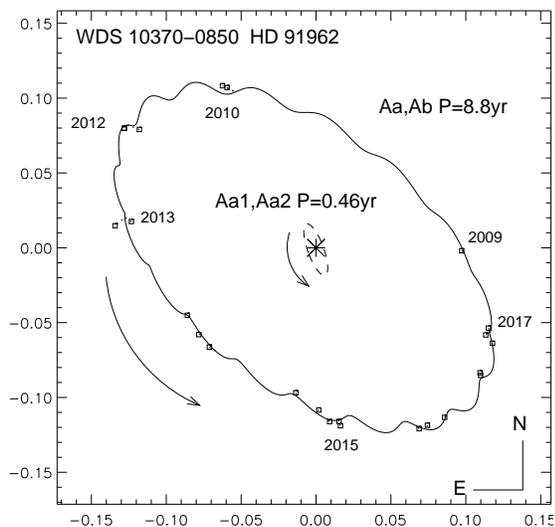}
\caption{The  observed motion  of the  subsystem TOK~44  Aa,Ab  in the
  quadruple star HD~91962  is modeled by two sets  of orbital elements
  with periods of 8.8 and 0.47 years (full line). The measurements are
  plotted as squares  and connected to the ephemeris  positions by the
  short dotted  lines.  The orbit of the  inner (unresolved) subsystem
  Aa1,Aa2 is shown by the dashed line; it causes the ``wobble'' in the
  observed motion of Aa,Ab. The scale is in arcseconds.
\label{fig:TOK44}
}
\end{figure} 

To illustrate the potential  of accurate HRCam measurements, the orbit
of  the  subsystem  TOK~44  Aa,Ab  in the  quadruple  system  HD~91962
\citep{planetary}   is  displayed  in   Figure~\ref{fig:TOK44}.   This
subsystem  was first resolved  at SOAR  in 2009  and now  the measures
cover almost one  full orbital period of 8.8  years.  All measurements
but one  are from HRCam.  The  orbit is a combined  solution that uses
radial  velocities  (RVs)  and   includes  the  motion  in  the  inner
(unresolved) subsystem  Aa1,Aa2 with the  period of 0.47 year  and the
estimated semimajor axis  of 18.4\,mas.  The mass of  the component Aa2
is  0.3 ${\cal M}_\odot$,  much smaller  than the  mass of  Aa1, 1.14
${\cal M}_\odot$, preventing the direct resolution of the inner pair at SOAR.
However, its  motion produces a detectable ``wobble''  in the relative
position of  Aa,Ab.  The wobble was  included in the  orbital model by
fitting the orientation, inclination, and astrometric axis $\alpha$ of
the inner orbit together with the positional measurements of Aa,Ab and
the RVs  of Aa.  The  derived elements of  the inner pair  Aa1,Aa2 are
$\alpha = 4.2  \pm 0.8$ mas, $\Omega = 21\degr \pm  11\degr$, and $i =
73\degr \pm  12\degr$. The amplitude of the  wobble, $\alpha$, matches
its  estimate  given in  \citep{planetary}.  The relative  inclination
between two  orbits computed from the  new elements is  small, $\Phi =
32\degr \pm  12\degr$, justifying the assumption  of orbit coplanarity
made in that paper.  The rms residuals of Aa,Ab in two coordinates are
2.2  and  2.1 mas.   If  the  wobble amplitude  is  set  to zero,  the
residuals increase  to 2.9 and 2.7  mas. This object is  a triple star
(the  outer  pair  A,B   is  resolved  at  0\farcs93).   Despite  this
complication,  the  measurements  of  the  inner  subsystem  are  very
accurate. Modeling of the wobble in another triple system, HIP~103987,
leaves the rms residuals of only 1.5 and 1.8 mas \citep{TL17}.

\subsection{Surveys}
\label{sec:survey}

The large  number of objects  that can be  observed with HRCam  in one
night  favor  its use  for    multiplicity  surveys. The  first
discoveries  of  48  close  binaries and  subsystems  were  unexpected
\citep{TMH10}. Later, known visual  binaries in the solar neighborhood
were  observed  systematically to  constrain  the  frequency of  inner
subsystems  \citep{Tok2014b}. This  effort has  helped to  improve the
multiplicity statistics \citep{Tok2014a}. Some of those subsystems how
have computed orbits (e.g. Figure~\ref{fig:TOK44}). 

The  HRCam  was   used  to  survey  75  {\it   Kepler}  objects  for
multiplicity \citep{Kepler},  although the bulk of such  work has been
done so far by other teams. As most of those stars are fainter than $I
= 12$ mag, the sensitivity was improved by closing the AO loop. The
same strategy was used in the multiplicity survey of young stars in
the Ori OB1 association, conducted in 2016 (Petr-Gotzens et al., in
preparation). As part of the same program, ten new young binaries in the
$\epsilon$~Cha association were discovered by \citet{BT17}. It is
noteworthy that 47 stars of this program were observed in 3.2 hours of
telescope time.


\section{Summary and outlook}
\label{sec:sum}

Systematic use  of the  new speckle camera  at the SOAR  telescope has
started  in 2008  and  continues at  present.   Thousands of  accurate
measurements of binary stars  delivered by this instrument have allowed
substantial improvement  of hundreds of visual  orbits and calculation
of many new  orbits. The impact of this data set  will extend far into
the future. More than 200 close binaries or subsystems were discovered
with  HRCam, contributing to the improved  statistics of  binary and
multiple stars. 

HRCam was  recently equipped with  the new, more  performant detector,
with a  better magnitude limit which  can be further  boosted by using
the  SAM  laser AO  system.   The high  efficiency  of  HRCam at  SOAR
(hundreds of  stars per  night) makes it  an instrument of  choice for
large surveys, surpassing the  capabilities of the existing AO systems
on large telescopes.  It will be an ideal instrument for the follow-up
of the  TESS targets, all brighter  than $I \approx 12$  mag. With the
current  efficiency,  the expected  3000  exoplanet host  candidates to  be
discovered by  TESS can be observed  in 10 nights.  In  a survey mode,
the  productivity of  HRCam observations  can be  further  improved by
using fixed instrument configuration and automating the star centering
and data taking.

Many  HRCam targets  are  relatively bright,  with  enough photons  to
correct the  wavefront in  real time. This  operational mode  has been
demonstrated in  2009--2010 during commissioning  of SAM \citep{SAM09},
but this capability is lost  now because SAM is permanently configured
for the  UV laser. If  a small dedicated  AO system with a  narrow FoV
were built,  it would greatly  benefit the speckle program.   It could
use one  region of the  spectrum (e.g.  green) for  wave-front sensing
while the remaining  wavelengths would go to the  HRCam detector.  The
concept of such instrument has been proposed by \citet{HARI}.  It will
be installed at the presently  unused side port of SOAR, without image
rotation. For faint targets, the  AO compensation will be partial, but
still useful, correcting in  real time focus and low-order aberrations
to get  the smallest possible  $\beta$.  Brighter stars will  be fully
compensated, allowing long exposures  for such challenging programs as
high-contrast  imaging   and/or  resolved  spectroscopy.    A  similar
combination  of AO  and SI/LI  is implemented  in the  AOLI instrument
\citep{AOLI}.








\acknowledgments  The   software  of  HRCam  has   been  developed  by
R.~Cantarutti  who modified  it  as necessary  to  adapt to  different
cameras. I am grateful to G.~Cecil and J.~Bispo for offering  their
Luca-R cameras at  times when our own camera  failed.  Special thanks to
N.~Law for loaning  his iXon-888 camera for the  speckle work at SOAR.
Comments  on this  paper by  S.~Hippler and  R.~Mendez has  helped the 
author to improve it. Detailed comments provided by the anonymous
Referee are also helpful.




\begin{thebibliography}{}






\bibitem[Baranec et al.(2014)]{Baranec2014}
Baranec, C., Riddle, R., Law, N. M. et al. 2014, ApJ, 790, 8


\bibitem[Brice\~no \& Tokovinin(2017)]{BT17}
Brice\~no, C. \& Tokovinin, A.  2017, AJ, 154, 195

\bibitem[Christou et al.(1985)]{Christou1985}
Christou, J. C., Cheng, A. Y. S., Hege, E. K. \& Roddier, C. 1985, AJ, 90, 2644

\bibitem[Gomez et al.(2016)]{Gomez2016}
Gomez, J., Docobo, J. A., Campo, P. P., \& Mendez, R. A.  2016, AJ, 152, 216

\bibitem[Hartkopf, Mason \& Worley (2001)]{VB6} 
Hartkopf, W. I., Mason, B. D. \& Worley, C. E. 2001, AJ, 122, 3472 (VB6)


\bibitem[Hartkopf et al.(2012)]{Hrt2012}
Hartkopf, W. I., Tokovinin, A., \& Mason, B. D.  2012, AJ, 143, 42 

\bibitem[Hippler et al.(2009)]{Hippler2009}
Hippler, S., Bergfors, C., Brandner W. et al. 2009, Messenger, 137, 14

\bibitem[Horch et al.(2012)]{Horch2012}
Horch, E. P., Howell, S. B., Everett, M. E. \& Ciardi, D. R. 2012, AJ,
144, 165

\bibitem[Hormuth et al.(2008)]{AstraLux}
Hormuth, F.,  Hippler, S., Brandner, W.  et al. 2008, Proc. SPIE, 7014, 48



\bibitem[Labeyrie(1970)]{Lab1970}
Labeyrie, A. 1970, A\&A, 6, 85

\bibitem[Law et al.(2016)]{HARI}
Law, N. M., Ziegler, C., \& Tokovinin, A.  2016, Proc. SPIE, 9907, id. 99070K

\bibitem[Maksimov et al.(2009)]{BTA}
Maksimov, A. F., Balega, Yu. Yu., Dyachenko, V. V. et al. 2009, AstBull, 64, 296

\bibitem[Mason et al.(2001)]{WDS}
Mason, B. D., Wycoff, G. L., Hartkopf, W. I. et al.  2001, AJ, 122, 3466 (WDS)


\bibitem[Mason et al.(2010)]{Msn2010}
Mason B. D., Hartkopf W. I., \& Tokovinin A.  2010, AJ, 140, 735-743.

\bibitem[McAlister et al.(1990)]{McAlister1990}
McAlister, H., Hartkopf, W. I. \& Franz, O. G. 1990, AJ, 99,  965

\bibitem[Mendez et al.(2017)]{Mdz17}
Mendez, R. A., Claveria, R. M., Orchad, M. E., \& Silva, J. F.  2017, AJ, 154, 187

\bibitem[Riddle et al.(2015)]{Riddle2015}
Riddle, R. L., Tokovinin, A., Mason, B. D. et al. 2015, ApJ, 799, 4

\bibitem[Schmitt et al.(2016)]{Kepler}
Schmitt, J. R., Tokovinin, A., Wang, Ji et al. 2016, AJ, 151, 159

\bibitem[Sebring et al.(2002)]{SOAR}
 Sebring, T. A., Krabbendam, V. L.  \&  Heathcote, S. 2002, Proc. SPIE, 4837, 71

\bibitem[Tighe et al.(2016)]{ADC}
Tighe, R., Tokovinin, A., Schurter, P. et al. 2016,  Proc. SPIE, 9908, 99083B

\bibitem[Tokovinin \& Cantarutti(2008)]{HRCam}
Tokovinin, A. \& Cantarutti, R. 2008, PASP 120, 170

\bibitem[Tokovinin et al.(2010a)]{TMH10}
Tokovinin A., Mason B. D., Hartkopf W. I. 2010a, AJ, 139, 743

\bibitem[Tokovinin et al.(2010b)]{SAM09}
Tokovinin A. Cantarutti R., Tighe R.,  2010b, PASP, 122, 1483


\bibitem[Tokovinin et al.(2014)]{TGM14}
Tokovinin, A. Gorynya, N.A. \& Morrell, N. I.  2014, MNRAS, 443, 3082

\bibitem[Tokovinin et al.(2015a)]{planetary}
Tokovinin, A., Latham, D. W. \& Mason, B. D.  2015a, AJ, 149, 195 

\bibitem[Tokovinin et al.(2015b)]{SOAR14}
Tokovinin, A., Mason, B. D., Hartkopf, W. I. et al.  2015b, AJ, 150, 50 

\bibitem[Tokovinin et al.(2016a)]{SOAR15}
Tokovinin, A., Mason, B. D., Hartkopf, W. I.  et al.  2016a, AJ, 151, 153

\bibitem[Tokovinin et al.(2016b)]{SAM}
Tokovinin, A., Cantaritti, R., Tighe, R. et al. 2016b,   PASP, 128, 125003

\bibitem[Tokovinin \& Latham(2017)]{TL17}
Tokovinin, A. \& Latham, D. W.  2017, ApJ, 838, 54

\bibitem[Tokovinin et al.(2018)]{SOAR16}
Tokovinin, A., Mason, B. D., Hartkopf, W. I.  et al.  2018, AJ, in preparation

\bibitem[Tokovinin(2012)]{Tok2012}
Tokovinin, A.  2012, AJ, 144, 56 


\bibitem[Tokovinin(2014a)]{Tok2014a}
Tokovinin, A. 2014a,  AJ,  147, 86 

\bibitem[Tokovinin(2014b)]{Tok2014b}  
Tokovinin,  A. 2014b, AJ, 148, 72

\bibitem[Tokovinin(2016a)]{Tok2016a}
Tokovinin, A.  2016a, AJ, 152, 10

\bibitem[Tokovinin(2016b)]{Tok2016b}
Tokovinin, A.  2016b, AJ, 152, 11

\bibitem[Tokovinin(2016c)]{Tok2016c}
Tokovinin, A.  2016c, AJ, 152, 138

\bibitem[Tokovinin(2016d)]{Tok2016d}
Tokovinin, A.  2016d, ApJ, 831, 151

\bibitem[Tokovinin(2017)]{Tok2017}
Tokovinin, A.  2017, AJ, 154, 110

\bibitem[Velaso et al.(2016)]{AOLI}
Velasco, S., Rebolo, R., Oscoz, A. et al. 2016, MNRAS, 460, 3519




\end{thebibliography}
\end{document}